\documentclass[pra,twocolumn]{revtex4}
\usepackage{graphics}
\usepackage{amsmath,amssymb}
\begin{document}
\title{Effective spin model for interband transport in a Wannier-Stark lattice system}

\author{Patrick Pl\"otz$^1$, Peter Schlagheck$^2$, and Sandro Wimberger$^1$}

\affiliation{$^1$Institut f\"ur Theoretische Physik, Universit\"at Heidelberg, Philosophenweg 19, 69120 Heidelberg, Germany\\ $^2$D\'{e}partement de Physique, Universit\'{e} de Li\`{e}ge, 4000 Li\`{e}ge, Belgium}
\date{\today}

\begin{abstract}
We show that the interband dynamics in a tilted two-band Bose--Hubbard model can be reduced to an analytically accessible spin model in the case of resonant interband oscillations. This allows us to predict the revival time of these oscillations which decay and revive due to inter-particle interactions. The presented mapping onto the spin model and the so achieved reduction of complexity has interesting perspectives for future studies of many-body systems.
\end{abstract}


\maketitle

\section{Introduction}\label{intro}

An amazing control of quantum degrees of freedom is nowadays routinely possible with the techniques of preparing and handling ultracold matter in the laboratory \cite{OberthalerReview,RevModPhys.79.235,BlochZwergerReview,RevModPhys.80.1215,RevModPhys.82.1225}. Backed by a plethora of theoretical proposals (see, e.g., \cite{Zoller,Diehl,Singh09,Verstraete}), a new direction is the coupling of such matter to additional degrees of freedom, such as provided by internal states (e.g. \cite{oberthaler-II}), by external potentials (e.g. \cite{oberthaler-I}), by coupling to a bath (e.g. \cite{Schmiedy-II}), to a continuum (e.g. \cite{Sias,Ghazal}) or even to macroscopic objects (e.g. \cite{PhysRevLett.101.183006,PhysRevLett.104.143002}). Such hybrid quantum systems are of high interest for applications, ranging from fundamental physics to metrology.

A major challenge in studying these systems is to reduce their inherent complexity. This is important for an understanding of both the internal dynamics as well as an extension to include a coupling to further degrees of freedom.
In this paper we focus on the dynamics of atomic bosons in a two band Bose-Hubbard model. The problem is non-stationary due to an additional Stark force (or constant tilt). In particular, we describe the Rabi-like oscillations between the two bands, which are well pronounced in the case of single-particle resonant tunnelling between the levels of adjacent lattice wells \cite{WSreview,Sias}. The presence of a second band gives an additional degree of freedom -- in the sense of the previous paragraph -- making the full many-particle problem very rich in new phenomena, yet also very complicated in general. We show how to effectively map the original problem to a much simpler spin system for specific fillings and parameters. This new model allows us to derive an analytical formula for the revival time of the interband oscillations which decay and revive due to weak inter-particle interactions.

\section{The system}\label{sec:system}
\subsection{The many-body model}\label{sec:manybodymodel}
We consider a two-band Bose-Hubbard model with an additional external force as obtained from a general many-body Hamiltonian under the assumption of a contact interaction and introduced in~\cite{TMW1,andrea}. We measure all parameters in recoil energies $E_{\rm rec}\equiv \hbar^2k_L^2/(2m)$, where $k_L$ is the wave vector of the laser creating the optical lattice and $m$ the mass of the atoms. Setting $\hbar = 1$ throughout, the Hamiltonian reads~\cite{TMW1,andrea}
\begin{align}\label{eq:fullHamiltonian}
\mathcal H & = \sum_{l=1}^L \Big[\epsilon_l^- n_l^a - \frac{t_a}{2}(a_{l+1}^{\dag}a_l^{} + { \rm h.c.}) + \frac{gW_a}{2} n_{l}^a (n_l^a-1)\notag   \\
& + \epsilon_l^+ n_l^b + \frac{t_b}{2}(b_{l+1}^{\dag}b_l^{} + { \rm h.c.}) + \frac{gW_b}{2} n_{l}^b (n_l^b-1) \notag  \\
& + FC_0(b_l^{\dag}a_l +  {\rm h.c.})+ 2gW_x n_{l}^a n_l^b + \notag \\ 
& + \frac{gW_x}{2}(b_l^{\dag}b_l^{\dag}a_l a_l + {\rm h.c.} ) \Big]. 
\end{align}
The operator $a_l$ ($a_l^{\dag}$) annihilates (creates) a particle at site $l$ of totally $L$ sites in the lower band and $b_l$ ($b_l^{\dag}$) in the upper band with the number operators $n_l^a = a_l^{\dag}a_l^{}$, $n_l^b = b_l^{\dag}b_l^{}$. The bands are separated by a bandgap $\Delta$ and have on-site energies $\epsilon_l^{\pm} = \pm\Delta/2+lF$. The hopping amplitudes between neighbouring sites in band $a, b$ are denoted by $t_a,t_b>0$, and a repulsive interaction between particles occupying the same site in band $a\;(b)$ with a strength $W_a\;(W_b)$ has been included. The single-particle coupling of the bands is proportional to the external Stark force $F$ via $C_0 F$ with a coupling constant $C_0$ depending on the depth of the lattice $V_0$~\cite{TMW1,andrea}. The bands are additionally coupled via the inter-particle interactions with a strength $W_x$. Focusing on a realisation with a single optical lattice, the parameters fulfill generally: $\Delta\gg t_a, t_b$, as well as $t_a,t_b\approx W_i$ and $C_0\approx -0.1$. We take the external force $F$ as a free parameter. Additionally we assume that the interaction strength can be tuned, e.g., by the use of Feshbach resonances~\cite{BlochZwergerReview}, and include a factor $g$ to all interaction terms. 
For numerical simulations, we change to the interaction-picture with respect to the external force~\cite{FB} which removes the tilt $\sum_l ln_l^{a,b}F$ and replaces $a_{l+1}^{\dagger}a_l^{}\rightarrow {\rm e}^{{\rm i} Ft}a_{l+1}^{\dagger}a_l^{}$ (and likewise for $b_{l+1}^{\dagger}b_l^{}$). The Hamiltonian is then time-dependent with a periodicity of $T_B\equiv 2\pi/F$ and allows to use periodic boundary conditions $a_{L+1} = a_1$ and $b_{L+1} = b_1$.

To study the interband transport, we prepare the system in an initial state $|\psi(0)\rangle$, with a uniform distribution of particles in the lower band only and evolve it in time by the many-body Schr\"odinger equation. The quantity we study is the (normalised) number of particles in the upper band
\begin{equation}\label{eq:occband2}
	\mathcal N_b (t) \equiv \frac{1}{N} \langle\psi (t)| \sum_l n_l^b |\psi(t)\rangle,
\end{equation}
where $N = \sum_l (n_l^a+n_l^b)$ is the total number of particles. We will refer to $\mathcal N_b(t)$ as \emph{occupation of the upper band}. For the range of parameters described above, this observable shows a superposition of many sinusoidal oscillations with an amplitude of few per cent, even for strong forces~\cite{collapse}.

\subsection{Weakly interacting system in resonance}\label{sec:noninter} 
Despite the small interband oscillations described in the previous paragraph, a strong enhancement of the interband transport is possible for specific parameter values. When the force-induced tilt of the lattice is such that a lower and upper band energy level become nearly degenerate (i.e.\ for $\Delta \approx m F,\; m\in \mathbb N$), the interband oscillations of $\mathcal N_b(t)$ come close to 100\% indicating a resonantly enhanced interband transport~\cite{collapse}. We refer to these specific parameter values as resonant and will focus on this resonant behaviour in the following. \\
To describe the non-interacting system $\mathcal H (g=0) = \mathcal H_0$ in resonance, it is useful to apply the following basis transformation involving Bessel functions of the first kind $J_n(x)$~\cite{collapse,Fukuyama} 
\begin{equation}\label{eq:basistransform}
\alpha_n  = \sum_{l\in\mathrm Z} J_{l-n}(x_a) a_l \qquad \beta_n  = \sum_{l\in\mathrm Z} J_{l-n}(x_b) b_l,
\end{equation}
with $x_{i} \equiv t_{i}/F$, $i = a,b$. By using $\sum_{l\in\mathrm Z} J_{n-l}(x)J_{n'-l}(x') = J_{n-n'}(x-x')$, one finds that the transformation removes the hopping terms from the original Hamiltonian~(\ref{eq:fullHamiltonian}) but leads to a coupling between any sites from the two different bands weighted by Bessel functions 
\begin{multline}\label{eq:transformedH}
\mathcal H_0 = \sum_{l\in\mathrm Z}\Big[ \epsilon_l^{-} \alpha_l^{\dagger}\!\alpha_l +\epsilon_l^{+}\beta_l^{\dagger}\!\beta_l  \\
	+ \sum_{m\in \mathrm Z} C_0F  J_{m}(\Delta x) ( \alpha_l^{\dagger}\!\beta_{l-m}  + \rm{h.c.} ) \Big],
\end{multline}
where $\Delta x = x_a + x_b$ and $\epsilon_l^{\pm} = \pm\Delta/2+lF$ as above. The resonance condition, $\Delta \approx mF$ or equivalently $\epsilon_{l}^{-}\approx\epsilon_{l-m}^{+} $, means that two levels from different bands become energetically degenerate. In this case, it is sufficient to keep only the direct coupling between these two sites leading to a sum of independent two-level systems
\begin{equation}\label{eq:resonantH}
	\mathcal H_0^{\rm res} = \sum_{l\in\mathrm Z}\Big[ \epsilon_l^{-} \alpha_l^{\dagger}\!\alpha_l +\epsilon_{l}^{+}\beta_{l}^{\dagger}\!\beta_{l}^{} 
	+ C_0F  J_{m}(\Delta x) ( \alpha_l^{\dagger}\!\beta_{l-m}  + \rm{h.c.} ) \Big].
\end{equation}
This approximate description of the system in resonance corresponds to lowest order nearly degenerate perturbation theory and higher order corrections are easily calculated, see e.g.~\cite{diss,twolevelsystem}. However, the lowest order approximation, eq.~(\ref{eq:resonantH}), gives already an accurate description of the non-interacting system in resonance~\cite{collapse,diss}. The resonant contribution to the non-interacting case is thus well-described as $\mathcal N_b(t)=\sin^2\left[C_0 F J_m(\Delta x)\, t\right]$ with a period $T_{\rm res} = \pi/[C_0 F J_m(\Delta x)]\gg T_B$ much larger than the Bloch period.

As demonstrated in~\cite{collapse}, the inclusion of a weak inter-particle interaction leads to a dephasing of the resonant interband oscillations. The occupation of the upper band as a function of time exhibits a collapse and revival effect, with the time-scales for the collapse and revival inversely proportional to the interaction strength $g$~\cite{diss}. An example of such oscillations under a weak repulsive interaction is given by the solid line in fig.~\ref{fig:spinmodel_occ} below. In the weakly interacting regime under consideration here, one of the interaction terms in the full Hamiltonian eq.~(\ref{eq:fullHamiltonian}) is most important. We focus solely on repulsive interactions, for which the system tries to avoid double occupancy of sites in either bands. However, the system is always assumed to be at approximately integer filling and at the same time sites from different bands are nearly degenerate, such that it cannot avoid to have two particles occupying the same site in either band. Thus the dominant contribution comes from the term $2gW_x\sum_ln_l^an_l^b$ (see~\cite{collapse} for further details). In the next section we derive an effective Hamiltonian that allows to study the effect of a weak interaction on the resonant interband oscillations in detail.

\section{Results}
\subsection{Effective spin model for system in resonance}
\begin{figure}
	\centering
	\resizebox{0.45\textwidth}{!}{\includegraphics{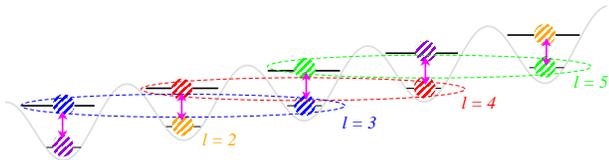}}
	\caption{Schematical view of the effect of the interband interaction on the system in resonance of order $m=2$. In resonance, the system is forced into a superposition of states from both bands (the sites forming a superposition are indicated by dashed ellipses). This happens on all lattice sites and the system cannot avoid an interaction of $2gW_x \sum_ln_l^an_l^b$. A new fictitious lattice labeling scheme setting the superpositions on one site is also indicated in the figure.}
	\label{fig:spinmodel}       
\end{figure}

The description of the non-interacting system in resonance according to eq.~(\ref{eq:resonantH}) contains already the seed for an effective model. The sum of many independent  two-level systems can be viewed as a system of non-interacting spins. We only need to re-order the labeling of lattice sites such that the two levels being coupled have the same site-index and the coupling operator is then proportional to the Pauli matrix $\sigma_x$. The constant of proportionality is the coupling matrix element $C_0FJ_m(\Delta x)$ from eq.~(\ref{eq:resonantH}). This is simply a different way of writing the approximate Hamiltonian for the non-interacting system in resonance, eq.~(\ref{eq:resonantH}), and is schematically displayed for a resonance of order $m=2$ in fig.~\ref{fig:spinmodel}.
To include the effect of the most important interaction term $2gW_x \sum_ln_l^an_l^b$ into our effective spin model, we insert the basis transformation from eq.~(\ref{eq:basistransform}) and obtain in the transformed basis
\begin{equation}\label{eq:interaction} 
\begin{split}
 & 2gW_x \sum_{l} n_l^a n_l^b=2gW_x \sum_{l,l_1,\ldots, l_4} \Big\{ J_{l-l_1}(x_a) J_{l-l_2}(x_a) \times \\ 
& \quad J_{l-l_3}(x_b) J_{l-l_4}(x_b) \; \alpha_{l-l_1}^{\dagger} \alpha_{l-l_2}^{} \beta_{l-l_3}^{\dagger} \beta_{l-l_4}^{} \Big\} \\
& \quad \approx  2gW_x J_0^2(x_a)J_0^2(x_b) \sum_{l} \alpha_{l}^{\dagger} \alpha_{l}^{} \beta_{l}^{\dagger} \beta_{l}^{} .
\end{split}\end{equation}
Here we used the fact that only one of the many different combinations of Bessel functions gives a significant contribution~\cite{collapse,diss}. The reason is that the arguments of the Bessel functions $x_a$ and $x_b$ are much smaller than unity for a realisation with a single optical lattice as discussed here, and the dominant contribution is therefore given by the product of four zeroth order Bessel functions $J_0^2(x_a) J_0^2(x_b)$. We denote the interaction strength for this dominant process by 
\begin{equation}\label{eq:definitionU}
 U \equiv 2 g W_xJ_0^{2}(x_a)J_0^{2}(x_b).
\end{equation}
The introduction of the new ficticious lattice is now effectively achieved by replacing $l\rightarrow l+m$ for the sites of the upper band. It is important to note that the interband interaction was between atoms occupying the same site in different bands in the original lattice, i.e.\, $\propto n_l^a n_l^b$, whereas in the new lattice it connects a particle at one lattice site in the lower band $\alpha_l^{\dagger}\alpha_l^{}\equiv n_l^{\alpha}$ with a particle at a different site in the upper band $\beta_{l+m}^{\dagger}\beta_{l+m}^{}\equiv n_{l+m}^{\beta}$ (where we used the transformed basis $\alpha_l,\beta_l$). We focus on unit filling $N=L$ and, since the repulsive interaction effectively suppresses higher occupation of lattice sites, we limit the occupation numbers of $n_l^{\alpha}$ and $n_l^{\beta}$ to $0$ or $1$ for our effective model. This allows us to replace $n_l^{\alpha, \beta}\rightarrow \sigma_l^{\uparrow,\downarrow}$ with the projectors on a spin-up or spin-down state $\sigma_l^{\uparrow\downarrow} =\left( 1_l \pm \sigma_l^z\right)/2$.

Collecting all arguments, the effective Hamiltonian (for a resonance of order $ m$) is accordingly given by
\begin{equation}\label{eq:spinHamil}
	\mathcal H_{\text{eff}} = \sum_{l=1}^{L} \left(V_m \sigma_l^x + U \sigma_l^{\uparrow} \sigma_{l+m}^{\downarrow}  \right)
\end{equation}
where $V_m= C_0FJ_m(\Delta x)$. Here $\sigma^{i}_l$ denotes the Pauli matrices for a spin at site $l$. The first part is as in the non-interacting resonant system, which was also a sum of independent two-level systems. We only changed the ordering of the sites to bring degenerate levels close together. The second part reflects the repulsion of two particles when sitting in different bands or different spin states respectively. Since we are using spin-$1/2$ matrices in this effective description it can only be applied to the case of unit filling and the number of lattice sites is per definition identical with the number of spins.  We expect it to be a good approximation for close-to-unit filling (as is supported by our results below, see fig.~\ref{fig:finitesizescaling}). An extension to higher fillings should be possible by using larger spins than spin 1/2, since this would allow further distinction of the type non-occupied, partly occupied, or highly-occupied, but is beyond the scope of the present article. The effective Hamiltonian (\ref{eq:spinHamil}) is translational invariant as our original model, such that one could use a reduction to subspaces of fixed total quasimomentum similar to~\cite{TMW1,diss,FB}. Please note, that the Hilbert space for the effective Hamiltonian has a dimension $\dim \mathcal H_{\rm eff}=2^L$ that is much smaller than the Hilbert space of the original bosonic problem, eq.~(\ref{eq:fullHamiltonian}), where $\dim \mathcal H = (N+2L-1)!/[N!(2L-1)!]$. This is advantageous for numerical computations since much larger system sizes become computable as compared to the original model. 

Let us discuss the effective model of eq.~(\ref{eq:spinHamil}) in more detail. The parameters in the effective spin model are chosen for the particular case of the system in resonance of order $m$. It includes only the resonant coupling between the two sites and other non-resonant couplings are neglected. The effective model does thus not reproduce small scale oscillations which are found on top of the resonant oscillations within the full model (see~\cite{collapse} for an example). However, these oscillations are only weakly influenced by the interparticle interaction and are not relevant for the collapse and revival effect we want to study. Another important aspect of the effective spin model concerns the choice of the interaction term $\sigma^{\uparrow}_l\sigma^{\downarrow}_{l+m}$. Here we included only one of the four interaction terms from the original Hamiltonian, eq.~(\ref{eq:fullHamiltonian}), which has the strongest effect on the resonant interband oscillations. The inclusion of the other terms is straightforward but not necessary in the present context. Furthermore we are limited to weak inter-particle interactions, $U\ll V_m$, which is, however, not a limitation of the effective model but originates from the physics of the original system: As soon as the inter-particle interaction becomes too strong, the resonant tunneling is washed out and the Rabi-like interband oscillations cede and eventually transform into an essentially structureless evolution of the band population defined in eq. \eqref{eq:occband2} \cite{diss}.

\begin{figure}
	\centering
	\resizebox{0.45\textwidth}{!}{\includegraphics{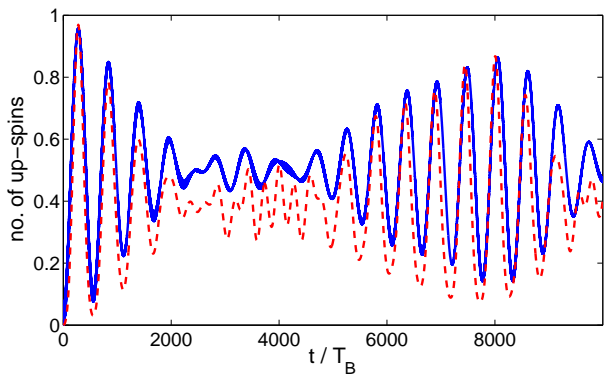}}
	\caption{Occupation of the upper band in the full two-band Bose--Hubbard model (solid line; see~\cite{collapse} for details) and number of up-spins in the effective spin model (dashed line) for the same parameters. The spin model reproduces most features of the time signal, first and foremost the collapse and revival are in very good agreement between both models. Parameters correspond to $V_0 = 10$: $\Delta = 7.77$, $t_a = 0.005$, $t_b = 0.121$, $C_0 = -0.114$, $W_a= 0.040$, $W_b=0.027$, and $W_x=0.018$; resonance order $m=1$, i.e. $F = 7.9804$; $g=0.1$ and $ N=5=L$.}
	\label{fig:spinmodel_occ}       
\end{figure}

To compare the effective model to the full problem, we computed the time-evolution of similar initial states in both models and show the resulting occupation of the upper band as a function of time with the pronounced collapse and revival effect in fig.~(\ref{fig:spinmodel_occ}). The occupation of the upper band for the full model is given by $\mathcal N_b(t)$, as defined by eq.~(\ref{eq:occband2}), and has been computed by direct numerical integration of the time-dependent Schr\"odinger equation. In the effective spin model, a state with an atom occupying the upper band is represented by a spin-up such that the corresponding observable for the spin model is given by $\mathcal{N}_{\uparrow}(t)= \tfrac{1}{L} \langle\psi(t)|\sum_l \sigma^{\uparrow}_l |\psi(t)\rangle$ and the initial state is of the form $|\!\!\downarrow\downarrow \ldots\downarrow\rangle$. Both observables are compared in fig.~\ref{fig:spinmodel_occ} for a weakly interacting system of medium size. 
Overall, the effective spin model reproduces the occupation of the upper band very well, especially when compared to the drastic simplification from the full two-band Bose--Hubbard model to the effective model of eq.~(\ref{eq:spinHamil}). The good agreement is particularly surprising when taking into account that a reduction of occupation numbers to 0 or 1 is usually known as ``hard-core bosons''~\cite{hardcore,hardcorelattice} and valid in the limit of strong interactions, whereas we are operating in exactly the opposite regime of $U\ll V_m$. Furthermore, the effective model reproduces the collapse and revival effect only when it is introduced from the transformed basis. Limiting the occupation number in the original $a_l,b_l$-basis by artificial constraints (such a truncation procedure was applied, e.g., in \cite{BHtruncated,TEBDJaksch}) cannot reproduce the effect~\cite{diss}. 

A great advantage of the effective spin model for the interband transport, eq.~(\ref{eq:spinHamil}), is its exact solvability. Rewriting the spin-up and -down operators in terms of Pauli matrices and applying a rotation of $\pi/2$ around the $y$-axis (which leads to $\sigma_x\rightarrow\sigma_z$ and $\sigma_z\rightarrow -\sigma_x$), our effective Hamiltonian takes the following form 
\begin{equation}\label{eq:spinny}
	\mathcal H_{\text{eff}} = \sum_{l=1}^{L} \left(V_m \sigma_l^x - \tfrac{1}{4}U\, \sigma_l^{z} \sigma_{l+m}^{z}  \right) + \text{const}.
\end{equation}
This Hamiltonian is known (for $m=1$) as the quantum Ising model in a transverse magnetic field~\cite{sachdev}. It describes coupled spins that tend to align in $ z$-direction but are subjected to the force of an applied magnetic field in $x$-direction. It can be solved exactly by subsequent application of a Jordan--Wigner transformation~\cite{JWTrafo}, Fourier and Bogolyubov transformation. 
The final result in terms of Bogolyubov quasi-particles is~\cite{diss,sachdev} 
\begin{equation}\label{eq:exactsolution}
 	\mathcal H_{\text{eff}} = \sum_{k}\epsilon(k)  \big( d_k^{\dagger} d_{k}^{} - 1/2\big). 
\end{equation}
The exact dispersion relation $\epsilon (k)$ is given by 
\begin{equation}\label{eq:dispersion}
\epsilon(k) = 2 V_m \sqrt{1- \tfrac{U}{2V_m}\cos k + (\tfrac{U}{4V_m})^{2}} \approx  2V_m - \tfrac{1}{2}U \cos k
\end{equation}
and can be approximated for our weakly interacting system $U\ll V_m$ as shown. Equation~(\ref{eq:exactsolution}) is the exact solution to our effective spin model. The elementary excitations of the system are non-interacting fermions with a dispersion relation that is approximately given by a cosine. These elementary excitations correspond to magnons, i.e., to delocalised spin-flips in the original spin basis. They read explicitly
\begin{gather}
 d_k  = \cos(\theta_k/2) c_k^{} - {\rm i} \sin(\theta_k/2) c_{-k}^{\dagger}\\
	\tan \theta_k = \frac{\sin k}{\cos k - 4 V_m/U},
\end{gather}
where $c_k$ is the Fourier transform of $c_l^{} = \sigma^{-}_{l} {\rm e}^{{\rm i} \pi \sum_{n<l} c_n^{\dagger}c_n^{}}$,
with $\sigma_l^{\pm}=(\sigma_l^{x}\pm{\rm i}\sigma_l^{y})/2$ and $\sigma^{z}_{l}  = 2c_l^{\dagger} c_l^{} -1$. 

\subsection{Revival time within the effective model}
\begin{figure}
	\centering
	\resizebox{0.475\textwidth}{!}{\includegraphics{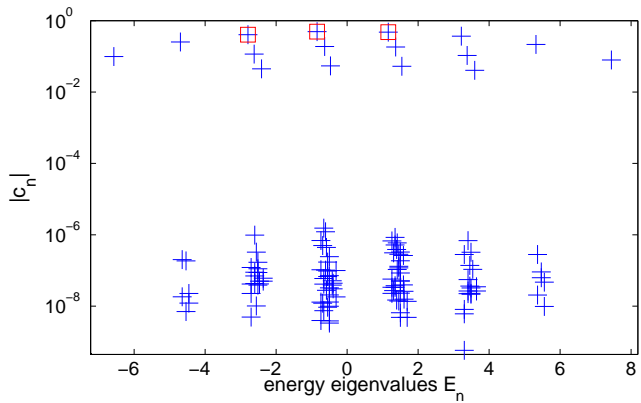}}
	\caption{Shown are the coefficients $c_n=\langle E_n| \psi (0)\rangle$ for an eigenbasis expansion of the initial state $|\psi(0)\rangle = |\!\!\downarrow \ldots\downarrow\rangle$ versus the corresponding eigenenergies ($+$). The three largest coefficients are marked by ($\square$) and highlight eigenstates with $M-1$, $M$, and $M+1$ magnons. Parameters: $V_m = 1$, $L=7$, $U=0.25$, and $F = 4.6020$. Note the logarithmic scale on the y-axis.}
	\label{fig:spinmodel_eigenvalues}       
\end{figure}

The exact solution eq.~(\ref{eq:exactsolution}) of the effective Hamiltonian allows, e.g., the computation of various correlation functions. But in the present context, we are interested in the time-evolution of particular initial states 
\begin{equation}
|\psi(t)\rangle = \sum_n {\rm e}^{-{\rm i}E_nt}\, c_n \,|E_n\rangle, 
\end{equation}
where $|E_n\rangle$ are the eigenstates to the effective model, eq.~(\ref{eq:exactsolution}), and $c_n=\langle E_n| \psi (0)\rangle$. In general the overlaps $c_n$ between the given initial state and all eigenstates are needed for the time evolution. 
Instead of an analytical derivation of the overlaps on basis of the Jordan--Wigner transformation, we adopt a numerical approach here in order to decide which of the magnon states are relevant. In detail, it is sufficient to know which eigenstates have a significant contribution to the time-evolution to estimate the revival time of the resonant interband oscillations. Fig.~\ref{fig:spinmodel_eigenvalues} shows the coefficients $c_n$ for a time evolution of the initial state $|\!\!\downarrow \ldots\downarrow\rangle$ sorted by their eigenenergies. 
The eigenenergies appear in several bunches corresponding to eigenstates with a different number of magnon excitations ranching from $0$ to $L$ magnons. Additionally, the three coefficients with the largest amplitude have been marked by squares in fig.~\ref{fig:spinmodel_eigenvalues}. We find that the largest coefficients in the eigenbasis expansion are from the energetically lowest eigenstates from the central bunches of the spectrum. To be more specific, we found numerically that the largest coefficients always come from the subspaces with $M-1$, $M$, and $M+1$ magnons, where $M = L/2$ for $L$ even and $M=(L-1)/2$ for $L$ odd, and are from the eigenstates with lowest energy within these subspaces. This important observation allows a simple estimate of the revival time as the time for a beating between oscillations with these three energies as frequencies.

\begin{figure}
	\centering
	\resizebox{0.45\textwidth}{!}{\includegraphics{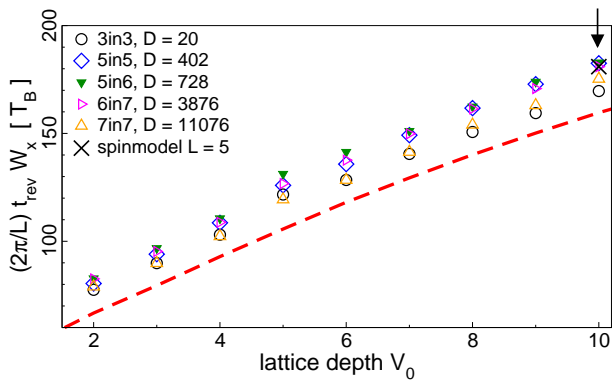}}
	\caption{We show the revival times from different numerical simulations of the full two-band Bose--Hubbard model (the corresponding system sizes `$N$in$L$' and the dimension of the Hilbert spaces $D$ are indicated in the legend). The numerical values have been rescaled by the size-dependent prefactor $2\pi/L$. The data points for different system sizes coincide, with small fluctuations due the approximation in eq.~(\ref{eq:energyshift}). Our analytical prediction eq.~(\ref{eq:trevspinmodel}) (red dashed line) captures nicely the scaling with the system parameters, yet it shows a systematic offset arising from the three level approximation. The numerical result of the full spin model (cross marked by arrow), extracted from fig.~\ref{fig:spinmodel_occ}, coincides with the full problem of eq.~\eqref{eq:fullHamiltonian}.}
	\label{fig:finitesizescaling}       
\end{figure}

To find the lowest eigenenergy of state with $M$ magnons, we use the fact that the energy of a many-body state with $M$ magnons in the weakly interacting regime is according to eq.~(\ref{eq:dispersion}) given by
\begin{equation}\label{eq:magnons}
E_M = \sum_{l = 1}^{M}\left(2V_m - \tfrac{1}{2}U \cos (k_{j_l})\right),
\end{equation}
where $k_{j_l} = 2\pi j_l/L$ and each $j_l$ can take a value between $1,\ldots,L$. The energies for a given number of magnons $M$ thus arise from different choices of the momenta $k_{j_l}$. The state with lowest energy in this cosine dispersion is obtained by using momenta that fill the empty cosine dispersion from zero upwards. A many-magnon state with $M=L/2$ is reached when half of the possible states are filled and with $M\pm1$ by adding or removing one magnon, respectively. This determines the momenta $k_{j_l}$ to obtain a state with $M$ magnons and minimal energy. 
We can now estimate the revival time from the difference between the energies of states with $M-1$, $M$, and $M+1$ magnons, i.e., we need 
$\Delta \omega = (E_{M+1}-E_{M}) - (E_{M} -E_{M-1} ) = E_{M+1} + E_{M-1} - 2 E_M. $
Inserting explicitly that the energy of $M$ magnons is proportional to the sum of $M$ cosine functions, eq.~(\ref{eq:magnons}), with different momenta filling the possible magnon states from below, we obtain the following frequency difference 
\begin{align}\label{eq:energyshift}
\Delta\omega & =  -\frac{U}{2} \Big( \sum_{j = 1}^{{M+1}} \cos (k_j) + \sum_{j=1}^{{M-1}} \cos (k_j)
	 -2 \sum_{j =1}^{M} \cos (k_j)\Big) \notag \\
	& =  -\frac{U}{2} \, \big(\cos k_{M+1} - \cos k_{M} \big) \approx  -\frac{\pi}{L}\;  U, 
\end{align}
where we expanded the cosine to lowest order around its zero. Using eq.~(\ref{eq:energyshift}) and our expression for $U$, eq.~(\ref{eq:definitionU}), we find that the revival time as estimated by oscillations between the dominant frequencies is given by
\begin{equation}\label{eq:trevspinmodel}
	\quad t_{\text{rev}} = \frac{L}{2\pi} \frac{4\pi}{g W_x J_0^2(x_a) J_0^2(x_b)}.
\end{equation}
The effective spin model predicts the revival time to be inversely proportional to the interaction strength and to a product of two Bessel functions from the basis transformation. The parameters of the original full Hamiltonian eq.~(\ref{eq:fullHamiltonian}), like the hopping strengths, the gap between the two energy bands and the order of the resonance, enter via the arguments of Bessel functions $x_{a,b} = t_{a,b}/F$, where the force has to be chosen according to the order of resonance $F\approx \Delta /m$. These parameters and the revival time change when the depth of the optical lattice $V_0$ is varied (see eq.~\eqref{fig:finitesizescaling}). Furthermore, the result eq.~(\ref{eq:trevspinmodel}) from the effective spin model additionally predicts a linear dependence of the revival time on the number of lattice sites. In this way, our effective model adds an additional factor of $L/2\pi$ to our earlier result~\cite{collapse}, which has been obtained following the arguments that led to eq.~\eqref{eq:definitionU} above, with the assumption that the initial state is comparable to a coherent state and estimating the revival time by computing the effect of the dominant interaction term on a coherent state perturbatively~\cite{collapse}.

To compare the result from eq.~(\ref{eq:trevspinmodel}) to numerical simulations of the full two-band Bose--Hubbard model, we use the size-dependent prefactor to rescale numerical results for different system sizes. The curves for different system sizes should coincide, as is verified in fig.~\ref{fig:finitesizescaling}. The revival times from full many-body models with Hilbert spaces ranging over three orders of magnitude fall onto one curve and demonstrate the validity of the effective spin model eq.~(\ref{eq:spinHamil}) for the interband transport in the weakly interacting two-band Bose--Hubbard model. The remaining fluctuations with the system size originate in the approximation of eq.~(\ref{eq:energyshift}) and decrease with growing $L$. The additional offset arises from taking only the three largest coefficients for the derivation of the explicit expression in eq.~(\ref{eq:trevspinmodel}). This leads to an underestimation by about 10\% of the time for the maximum in the revived interband oscillations (which is our definition for the revival time). Inclusion of more than three coefficients should remove this systematic offset between the predicted and measured revival time. This has been tested by comparing the oscillation dynamics of the original model \eqref{eq:fullHamiltonian} with the full spin model \eqref{eq:spinHamil} \cite{diss}, and the good agreement is shown for an exemplary data point in fig. \ref{fig:finitesizescaling} (cross, extracted from the temporal evolution presented in fig.~\ref{fig:spinmodel_occ} above).
We finally note, that eq.~(\ref{eq:trevspinmodel}) predicts a divergence of the revival time whenever the parameters of the system are chosen such that one of the Bessel functions in the denominator vanishes. This can be achieved, e.g., by tuning the energy gap $\Delta$ between the two bands, and the expected divergence of the revival time close to Bessel zeros was observed in numerical simulations~\cite{diss}, giving room for a great deal of control of the resonant interband oscillations.

\section{Summary}
We have shown how to reduce the complexity of the original Hamiltonian of eq.~\eqref{eq:fullHamiltonian} to the exactly solvable model of eq.~\eqref{eq:spinny} for filling factors of the order one and for resonant coupling between the two energy bands. For weak inter-particle interactions the model is in good agreement with the full problem and allowed us to derive an analytical formula for the revivals of the resonant interband oscillations.  Interesting future aspects to work on would be to include decay to higher energy states in the continuum part of the spectrum (e.g. by opening the model in a similar way as exercised in \cite{TMW1,TMW2} for a one-band problem) and to extend the problem to atoms with internal structure. The internal degrees of freedom would become correlated with the external transport in ``horizontal'' -- along the lattice -- and ``vertical'' --  between the bands -- direction. Reductions of complex models are in general a necessary prerequisite in order to describe quantum systems with many degrees of freedom -- possibly of different kind and nature. 
So we hope that the spirit of our approach may inspire future research in this direction.

\begin{acknowledgements}
This work was supported by the DFG FOR760. SW is especially grateful to the Hengstberger Foundation for the Klaus-Georg and Sigrid Hengstberger Prize. PP and SW acknowlegde furthermore support from the Klaus Tschira Foundation, by the Helmholtz Alliance Program of the Helmholtz Association (contract HA-216 ``Extremes of Density and Temperature: Cosmic Matter in the Laboratory''), and within the framework of the Excellence Initiative by the German Research Foundation (DFG) through the Heidelberg Graduate School of Fundamental Phys\-ics (grant number GSC 129/1), the Frontier Innovation Fonds and the Global Networks Mobility Measures.
\end{acknowledgements}

\bibliography{biblio}

\end{document}